\documentclass[12pt]{article}
\usepackage{graphicx}

\usepackage{hyperref}
\usepackage{textcomp}
\usepackage{color}

\makeatletter
\g@addto@macro\bfseries{\boldmath}
\makeatother

\newcommand\pubnumber{}
\newcommand\pubdate{November 15, 2014}

\def\heidelberg{
Physikalisches Institut, Im Neuenheimer Feld 226, \\
69120 Heidelberg, Germany}

\def\Title#1{\begin{center} {\Large #1 } \end{center}}
\def\Author#1{\begin{center}{ \sc #1} \end{center}}
\def\Address#1{\begin{center}{ \it #1} \end{center}}

\newcommand\pubblock{\rightline{\begin{tabular}{l} \pubnumber\\
         \pubdate  \end{tabular}}}
\newenvironment{Abstract}{\begin{quotation}  }{\end{quotation}}
\newenvironment{Presented}{\begin{quotation} \begin{center} 
             PRESENTED AT\end{center}\bigskip 
      \begin{center}\begin{large}}{\end{large}\end{center} \end{quotation}}
\def\Acknowledgements{\bigskip  \bigskip \begin{center} \begin{large}
             \bf ACKNOWLEDGEMENTS \end{large}\end{center}}

\def\beq{\begin{equation}}
\def\eeq#1{\label{#1}\end{equation}}
\def\eeqn{\end{equation}}

\def\beqa{\begin{eqnarray}}
\def\eeqa#1{\label{#1}\end{eqnarray}}
\def\eeqan{\end{eqnarray}}

\let\bar=\overbar

\def\Dslash{\not{\hbox{\kern-4pt $D$}}}
\def\dslash{\not{\hbox{\kern-2pt $\del$}}}

\def\msb{{\bar{\ssstyle M \kern -1pt S}}}

\begin{document}
\begin{titlepage}
\pubblock

\vfill
\Title{Lifetime measurements in $b$-hadron decays at LHCb}
\vfill
\Author{Francesca Dordei \\\vspace{0.2cm} {\rm on behalf of the LHCb Collaboration}}
\Address{\heidelberg}
\vfill
\begin{Abstract}
Precision lifetime measurements of $b$-flavoured hadrons are an important test of the validity of the theoretical tool used to determine $b$-hadrons observables, the Heavy Quark Expansion. Recent measurements of the $B^+$, $\Lambda^0_b$, $\Xi^-_b$, $\Xi^0_b$ and $\Omega^-_b$ hadrons lifetimes are reported. Moreover, several $B^0$ and $B^0_s$ effective lifetime measurements are discussed, as well as a measurement of the decay width difference in the $B^0$ system, $\Delta \Gamma_d$. All the measurements have been performed using $pp$ collision data collected with the LHCb detector.
\end{Abstract}
\vfill
\begin{Presented}
8th International Workshop on the CKM Unitarity Triangle (CKM 2014) \\ September 8-12, 2014, Vienna, Austria
\end{Presented}
\vfill
\end{titlepage}
\def\thefootnote{\fnsymbol{footnote}}
\setcounter{footnote}{0}

\section{Introduction}

Weak decays of hadrons that contain a $b$ quark, hereafter referred to as $b$ hadrons, are important to constraints the parameters coming from the flavour sector of the Standard Model. On the other hand, they represent also a good probe of that part of the
strong-interaction phenomenology that is least understood: the confinement of quarks
and gluons inside hadrons. 
In particular, precision measurements of $b$-hadron lifetimes are an important test of the theoretical approach to $b$-hadron observables, known as \textit{Heavy Quark Expansion} (HQE). In this approach the decay rate is calculated as an expansion in inverse powers of the heavy b-quark mass~\cite{Bigi,Uraltsev,Lenz}: 
\begin{equation}
\label{eq:lifetime}
\Gamma = \Gamma_0 + (\Lambda^2/m_b^2)\Gamma_2 + (\Lambda^3/m_b^3)\Gamma_3 + ... \,.
\end{equation}
Here $\Gamma_0$ represents the decay of a free $b$ quark, and according to this contribution all \mbox{$b$ hadrons} have the same lifetime.
Different corrections at order $\Lambda^i/m^i_b$ with $i=2,3,...$ due to the kinetic and the chromomagnetic operators, and to spectator quark(s) involved in the decay alter the lifetime at approximately the $10\%$ level.
The HQE model allows to determine the theoretical predictions of several key $b$-hadron observables that are used to test the validity of the Standard Model (SM). Thus, it is of vital important to confirm the validity of HQE predictions to a high accuracy. \\
  
Measurements of the so-called \textit{effective lifetime} in neutral $B$-meson decays allow to probe the decay width differences $\Delta \Gamma_s$ and $\Delta \Gamma_d$ of the $B^0_s$ and $B^0$ system, respectively. Moreover, the $C\!P$-violating phase $\phi_s$ of the $B^0_s-\overline{B}^0_s$ mixing box-diagram~\cite{Fleischer} can be extracted. 

\paragraph{Determination of $\Delta \Gamma_s$ and $\phi_s$} Consider a $B^0_s (\overline{B}^0_s) \rightarrow f$ transition, with a final state $f$ into which both a $B^0_s$ and a $\overline{B}^0_s$ meson can decay. When the initial flavour of the $B^0_s$ meson is unknown, the corresponding untagged decay rate can be written as follows\footnote{Natural units where $\hbar=c=1$ are used in these proceedings.}:
\begin{equation}
\label{eq:decayrate}
 \Gamma (B^0_s \ (\overline{B}^0_s) \ (t) \rightarrow f) = \mathcal{N}_f \left| A_f \right|^2  
         \, e^{- \Gamma_s t}  \left\{ 
        \cosh \frac{\Delta \Gamma_s \,  t}{2} +  \, 
        \mathcal{A}_{\Delta \Gamma_s} \, \sinh \frac{\Delta \Gamma_s \,  t}{2} 
        \right\} 
\end{equation}

Here, $\Delta \Gamma_s \equiv \Gamma_L - \Gamma_H$ is the decay width difference where the quantities $\Gamma_H$ and $\Gamma_L$ are the decay widths of the heavy and light $B^0_s$ mass eigenstates. The parameter $\mathcal{A}_{\Delta \Gamma_s}$ is defined as $\mathcal{A}_{\Delta \Gamma_s} = -2\Re e(\lambda)/(1+|\lambda|^2)$, where $\lambda \equiv (q/p)(\overline{A}/A)$. The complex coefficients $p$ and $q$ define the mass eigenstates of the $B^0_s-\overline{B}^0_s$ system in terms of the flavour eigenstates (see e.g. Ref.~\cite{LenzNierste}) and $A(\overline{A})$ is the amplitude for a $B^0_s(\overline{B}^0_s)$ to decay to $f$. In presence of $C\!P$ violation the parameter $\mathcal{A}_{\Delta \Gamma_s}$ is a function of $\phi_s$~\cite{Fleischer}.
The effective lifetime of the decay $B^0_s (\overline{B}^0_s) \rightarrow f$ is defined as the time expectation value of the untagged decay rate:
\begin{equation}
  \tau_f^{\rm eff} \equiv \frac{\int_0^\infty t \,\,\langle \Gamma (B^0_s(\overline{B}^0_s)(t) \rightarrow f) \rangle dt}{\int_0^\infty \langle \Gamma (B^0_s(\overline{B}^0_s)(t) \rightarrow f) \rangle dt}
\end{equation}
which is equivalent to the lifetime that results from fitting the untagged decay time distribution in Eq.~\ref{eq:decayrate} with a single exponential. It is a function of the final state $f$ as it depends on the relative proportions of the heavy and light contributions in a given decay.
By making use of the definition $y_s \equiv \Delta \Gamma_s / 2\Gamma_s$ and using $\tau_{B^0_s} \equiv \Gamma_s^{-1}$, the effective lifetime can be expressed as:
\begin{equation}
\tau_f^{\rm eff} = \tau_{B^0_s} + \tau_{B^0_s}\, \mathcal{A}_{\Delta \Gamma_s} y_s + \tau_{B^0_s} [2-\mathcal{A}_{\Delta \Gamma_s}^2] y_s^2 + \mathcal{O}(y_s^3).
\end{equation}
Thus, an effective lifetime measurement can be used to constrain $y_s$ (therefore $\Delta \Gamma_s$) with respect to $\phi_s$, with the advantage that only an untagged analysis is needed.

\paragraph{Determination of $\Delta \Gamma_d$}\label{sec:deltagamma}

The decay width difference in the $B^0$ system, $\Delta 
\Gamma_d$, is sufficiently small \cite{Lenz} that the heavy and light mass eigenstates cannot be resolved experimentally. It can be constrained using the effective $B^0$ lifetime measurements in a flavour specific decay, like $B^0\to J/\psi K^*(892)^0$, and in a decay to a $C\!P$ eigenstate, like $B^0\to J/\psi K^0_S$, as proposed in 
\cite{Gershon:2010wx}.
Defining the ratio, $R$, between the effective lifetimes in these two final states as
$ R \equiv \tau_{B^0 \to J/\psi K^0_{\rm{\scriptscriptstyle S}}}^{\mathrm{eff}}/\tau_{B^0 \to J/\psi K^*(892)^0}^{\mathrm{eff}} $,
the expression for $\Delta \Gamma_d/\Gamma_d$ is
\begin{equation}
 \frac{\Delta\Gamma_d}{\Gamma_d} = \frac{2}{\cos{(2 \beta)}}\, (R-1) + \frac{2}{\cos{(2 \beta)}^2}\, (R-1)^2 + \mathcal{O}((R-1)^3).
\end{equation}
Here $\beta$ is given by $\arg{[-(V^{\textcolor{white}{*}}_{cd}V^*_{cb})/(V^{\textcolor{white}{*}}_{td}V_{tb}^*)]}$, where $V_{ij}$ are elements of the Cabibbo-Kobayashi-Maskawa (CKM) matrix.

\section{Results}
The results presented here has been measured using $pp$ collision data corresponding to an integrated luminosity of $1$ (3) $\mathrm{fb}^{-1}$, collected with the LHCb detector~\cite{LHCb} at a centre-of-mass energies of $\sqrt{s}=7\ (7-8)$ TeV.
At LHCb, $b$ hadrons are produced with an average momentum of around $100$ $\mathrm{GeV}/\mathrm{c}$ and have decay vertices displaced, around 1-2 cm, from the primary $pp$ interaction vertex (PV). On the contrary, combinatorial background candidates, which corresponds to random combination of tracks, tend to have low momentum and to originate from the PV.
Conventional selections exploit these features to select $b$ hadrons by requiring that their decay products are energetic and significantly displaced from the primary interaction vertex. 
The distance of closest approach (impact parameter) of $b$-hadron decay products to any primary vertex or the distance traveled by the $b$ hadron are important discriminant variables. These requirements introduce efficiencies that depend on the $b$-hadron decay-time, collectively called decay-time acceptance, which need to be taken into account in the analyses.
Moreover, there could be also implicit biases, for example due to geometrical acceptance of the detector and time-dependent efficiencies introduced in the reconstruction process.
Experimentally, it is challenging to correct for these effects. Two strategies are commonly used to make an unbiased lifetime measurement. A so-called \textit{relative measurement}, where the $b$-hadron lifetime is measured relative to a well known lifetime with the effect that the decay-time acceptance mostly cancel or a so-called \textit{absolute measurement}, where the decay-time acceptance is investigated and corrected. Both these techniques have been used to perform the measurements reported in these proceedings.

\paragraph{$B^0_s \to K^+ K^-$ lifetime with $\mathcal{L}=1\ \mathrm{fb}^{-1}$}
A measurement of the effective lifetime of the $B^0_s \rightarrow K^+ K^-$ (charge conjugate modes are implied throughout) decay is of considerable interest as it is sensitive to new physics phenomena affecting the $B^0_s$ mixing phase and entering the decay at loop level. 
The $K^{+}K^{-}$ final state is mainly $C\!P$-even and in the Standard Model $\mathcal{A}_{\Delta \Gamma_s}$ is predicted to be equal to $\mathcal{A}_{\Delta \Gamma_s}^{\rm} = -0.972^{+0.014}_{-0.009}$~\cite{KK}. Hence, the effective lifetime is approximately equal to $1/\Gamma^{s}_L$ and it is predicted to be equal to $\tau^{\rm eff,\, SM}_{KK} = 1.395 \pm 0.020\, \mathrm{ps}$. 
The analysis presented here~\cite{KK} is a reanalysis of the 2011 data sample collected at LHCb using a data-driven method to determine a per-event correction for the decay-time acceptance. In this way, the full statistical power of the data set is exploited. Moreover, the effective lifetimes in the $B^0 \to K^+ \pi^-$ and $B^0_s \to \pi^+ K^-$ decays have also been measured.
The effective $B^0_s \rightarrow K^+ K^-$ lifetime is extracted using an unbinned maximum likelihood fit to the $B^0_s$ invariant mass and decay time distributions. First, a fit to the invariant mass distribution is perfomed, which finds $10471\pm121$ $B^0_s \rightarrow K^+ K^-$ candidates. It is used to determine the signal and background probabilities of each candidate~\cite{sweight}. By using the signal probability, the  $B^0_s \rightarrow K^+ K^-$ signal decay-time distribution is isolated. The effective $B^0_s \rightarrow K^+ K^-$ lifetime is found to be $\tau_{KK} = 1.407 \pm 0.016\, \mathrm{(stat.)} \pm 0.007\, \mathrm{(syst.)}$~$\mathrm{ps}$, in good agreement with the SM prediction and a previous independent measurement from LHCb~\cite{ind_lif}. This measurement is used to perform a first direct determination of the asymmetry parameter $\mathcal{A}_{\Delta \Gamma_s}$ for this decay mode, which gives $\mathcal{A}_{\Delta \Gamma_s} = -0.87 \pm 0.17\, \mathrm{(stat.)} \pm 0.13\, \mathrm{(syst.)}$. This value is consistent with the level of $C\!P$ violation predicted by the SM.

\paragraph{$\overline{B}^0_s \to D^+_s D^-_s$ lifetime with $\mathcal{L}=3\ \mathrm{fb}^{-1}$}\label{subsec:an6}
The measurement of the effective lifetime of the $\overline{B}^0_s$ meson in the decay \mbox{$\overline{B}^0_s \to D^+_s D^-_s$} is reported~\cite{lifetime}, using a dataset corresponding to $3\,\,\mathrm{fb}^{-1}$ of integrated luminosity. The $\overline{B}^0_s \to D^+_s D^-_s$ lifetime is measured relative to the well-known $B^-$ lifetime, using the decay \mbox{$B^- \to D^0 D^-_s$}, which has a similar topology and kinematic properties. 
The measured value of the $\overline{B}^0_s \to D^+_s D^-_s$ effective lifetime is $\tau = 1.379 \pm 0.026 \,\mathrm{(stat.)} \pm 0.017) \,\mathrm{(syst.)}$~ps. The $D^+_s D^-_s$ final state is $C\!P$-even hence the effective lifetime is approximately equal to $\Gamma^{-1}_L$. Using this fact a value of $\Gamma_L = 1.52 \pm 0.15 \,\mathrm{(stat.)} \pm 0.001 \,\mathrm{(syst.)}\,\,\mathrm{ps}^{-1}$ is derived.

\paragraph{$B^0_s \rightarrow D^-_s\, \pi^+$ lifetime with $\mathcal{L}=1\ \mathrm{fb}^{-1}$}

The $B^0_s \rightarrow D^-_s\, \pi^+$ is a decay to a so-called flavour specific final state. Equal proportions of heavy and light mass eigenstates contributes to this decay at $t=0$. The $B^0_s$ flavour specific effective lifetime can be approximated to
\begin{equation}
 \tau_{fs}^{\rm eff} \approx \frac{1}{\Gamma_s}\frac{1+ \left(\frac{\Delta \Gamma_s}{2 \Gamma_s}\right)^2}{1- \left(\frac{\Delta \Gamma_s}{2 \Gamma_s}\right)^2}\ .
\end{equation}
In the SM it is predicted to be $\tau_{fs}^{\rm eff,\, SM}= 1.009 \pm 0.004$.
The $B^0_s$ time-dependent decay rate is measured \cite{Dminus} with respect to the well measured lifetime of the $B^+$ and $B^0$ mesons, which are reconstructed in three different final states with similar topology and kinematic properties. Measuring the $B^0_s$ lifetime with respect to the three normalisation channels gives three consistent results which are fully correlated. Thus, the one with the smallest overall uncertainty is chosen. The $B^0_s$ flavour specific effective lifetime is measured to be equal to $\tau_{fs}^{\rm eff} = 1.535 \pm 0.015\, \mathrm{(stat.)} \pm 0.012\, \mathrm{(syst.)} \pm 0.007 \, \mathrm{(input)}$ $\mathrm{ps}$. The third uncertainty is due to the input decay lifetime of the normalisation channel \cite{pdg}. The measurement is in good agreement with the SM prediction.

\paragraph{$H_b \rightarrow J/\psi\,\, X$ lifetimes with $\mathcal{L}=1\ \mathrm{fb}^{-1}$}

Absolute measurements of the $B^+ \to J/\psi K^+$, $B^0 \to J/\psi K^*(892)^0$, $B^0 \to J/\psi K^0_S$, $\Lambda^0_b \to J/\psi \Lambda$ and $B^0_s \to J/\psi \phi$  lifetimes have been performed \cite{hb} using a data sample corresponding to an integrated luminosity of 1\,fb$^{-1}$. Collectively, all these modes are referred to as $H_b \rightarrow J/\psi\,\, X$. All the different channels have a $J/\psi$ in the final state and are triggered and selected in a uniform way. Efficiencies that depend on the decay-time are introduced particularly during the reconstruction of the tracks associated to the final state particles. A data-driven technique has been developed to determine these efficiencies. After an event-by-event correction, the lifetimes of the different $b$ hadrons are determined from an unbinned maximum likelihood fit to the $b$-hadron invariant mass and decay-time distributions. The measured $b$-hadron lifetimes are reported in Table~\ref{tab:sfit_fit_results}.
All results are compatible
with existing world averages~\cite{HFAG}. 
With the exception of the $\Lambda^0_b \to J/\psi \Lambda$  channel, these are the single most precise measurements of the $b$-hadron lifetimes.

\begin{table}[t]
\caption{\small Fit results for the $B^+$, $B^0$, $B^0_s$ mesons and $\Lambda^0_b$ baryon lifetimes. The first uncertainty is
statistical and the second is systematic.}
\centerline{
\begin{tabular}{lc}
        \hline \hline
	Lifetime		&	Value [ps]\\
	\hline
	$\tau_{B^+ \to J/\psi K^+}$	  &	1.637 $\pm$ 0.004 $\pm$ 0.003 \\
	$\tau_{B^0 \to J/\psi K^*(892)^0}$&	1.524 $\pm$ 0.006 $\pm$ 0.004 \\
	$\tau_{B^0 \to J/\psi K^0_S}	$&	1.499 $\pm$ 0.013 $\pm$ 0.005 \\
	$\tau_{\Lambda^0_b \to J/\psi \Lambda}$	  &	1.415 $\pm$ 0.027 $\pm$ 0.006 \\
	$\tau_{B^0_s \to J/\psi \phi}	$&	1.480 $\pm$ 0.011 $\pm$ 0.005 \\
	\hline
\end{tabular}
}

\label{tab:sfit_fit_results}
\end{table}

In order to better compare these results with the HQE predictions, several lifetime ratios have been determined.
Table~\ref{tab:ratio_results} reports the ratios of the $B^+$, $B^0_s$ and $\Lambda^0_b$ lifetimes
to the $B^0$ lifetime measured in the flavour-specific $B^0 \to J/\psi K^*(892)^0$ channel. 
All ratios are consistent with SM
predictions~\cite{Lenz}
and with previous measurements~\cite{HFAG}.
Furthermore, the ratios $\tau_{B^+}/\tau_{B^-}$, $\tau_{\Lambda^0_b}/\tau_{\overline{\Lambda}^0_b}$ and
$\tau_{B^0 \to J/\psi K^*(892)^0}/\tau_{\overline{B}^0\to J/\psi \overline{K}^{*}(892)^0}$
are reported. Measuring any of these different from unity would indicate a violation of $C\!PT$ invariance or, for
\mbox{$B^0 \to J/\psi K^*(892)^0$} decays, could also indicate that $\Delta\Gamma_d$ is 
non-zero and $B^0 \to J/\psi K^*(892)^0$ is not $100\%$ flavour-specific.
No deviation from unity of these ratios is observed. 
\begin{table}[t]
\caption{\small Lifetime ratios for the $B^+$, $B^0$, $B^0_s$ mesons and $\Lambda^0_b$ baryon. The first uncertainty is
statistical and the second is systematic.}
\centerline{
\begin{tabular}{lc}
        \hline \hline
	Ratio			&	Value\\
	\hline
	$\tau_{B^+}/\tau_{B^0 \to J/\psi K^*(892)^0}$		&	1.074 $\pm$ 0.005 $\pm$ 0.003 \\
	$\tau_{B^0_s}/\tau_{B^0 \to J/\psi K^*(892)^0}$	&	0.971 $\pm$ 0.009 $\pm$ 0.004 \\
	$\tau_{\Lambda^0_b}/\tau_{B^0 \to J/\psi K^*(892)^0}$	&	0.929 $\pm$ 0.018 $\pm$ 0.004 \\\hline
	$\tau_{B^+}/\tau_{B^-}$		&	1.002 $\pm$ 0.004 $\pm$ 0.002 \\
	$\tau_{\Lambda^0_b}/\tau_{\overline{\Lambda}^0_b}$	&	0.940 $\pm$ 0.035 $\pm$ 0.006 \\
	$\tau_{B^0 \to J/\psi K^*(892)^0}/\tau_{\overline{B}^0\to J/\psi \overline{K}^{*}(892)^0}$		&	1.000 $\pm$ 0.008 $\pm$ 0.009 \\
	\hline
\end{tabular}
}
\label{tab:ratio_results}
\end{table}
The effective lifetimes of \mbox{$B^0 \to J/\psi K^*(892)^0$} and $B^0 \to J/\psi K^0_S$ decays are used to
measure $\Delta\Gamma_d/\Gamma_d$ as introduced in Section \ref{sec:deltagamma}. 
Using the effective lifetimes reported in Table~\ref{tab:sfit_fit_results} and
$\beta = (21.5^{+0.8}_{-0.7})^{\circ}$~\cite{HFAG}, $\Delta\Gamma_d/\Gamma_d$ is measured to be
$\Delta \Gamma_d/\Gamma_d = -0.044 \pm 0.025 \, \mathrm{(stat.)}  \pm 0.011\, \mathrm{(syst.)} $, consistent with the SM expectation and
the current world-average value~\cite{HFAG}.

\paragraph{$\Lambda^0_b \rightarrow J/\psi\,\, p\,\, K^-$ lifetime with $\mathcal{L}=3\ \mathrm{fb}^{-1}$}

Historically, discrepancies between the theoretical predictions and the experimental measurements of the $\Lambda^0_b$ lifetime have questioned the validity of HQE. 
The LHCb collaboration performed a measurement of this lifetime ratio utilizing the
\mbox{$\Lambda^0_b \rightarrow J/\psi p K^-$} decay using \mbox{$\mathcal{L}=1\ \mathrm{fb}^{-1}$} of data \cite{lambda1}. This $\Lambda^0_b$ decay mode was
first seen by LHCb. It has a better reconstruction efficiency than the $J/\psi \Lambda$
final state \cite{hb}, as in the latter one the $\Lambda$ is a long-lived particle and most of the decay products do not traverse the whole LHCb tracking system. The lifetime ratio measurement has been updated using a sample corresponding to $\mathcal{L}=3\ \mathrm{fb}^{-1}$. It is a relative measurement, where the topologically similar normalisation channel chosen is $B^0 \to J/\psi K^- \pi^+$. An unbinned maximum likelihood fit to the $\Lambda^0_b$ and $B^0$ invariant mass distributions gives $50233 \pm 331$ $\Lambda^0_b$ signal events.
From a fit to the ratio of the background subtracted decay-time distributions in these two channels, the ratio of lifetimes is measured to be $\tau(\Lambda^0_b)/\tau(B^0) = 0.974 \pm 0.006\, \mathrm{(stat.)} \pm 0.004\, \mathrm{(syst.)}$. Multiplying the lifetime ratio by $\tau(B^0 ) =
1.519 \pm 0.007$ ps \cite{pdg}, the $\Lambda^0_b$ baryon lifetime is determined to be $\tau(\Lambda^0_b) = 1.479 \pm 0.009\, \mathrm{(stat.)} \pm 0.010\, \mathrm{(syst.)}$ ps. This is the most precise measurement to date and it demonstrates that the $\Lambda^0_b$ lifetime is shorter than the $B^0$ lifetime by $-(2.7 \pm 0.7)\%$, consistent with the predictions of the HQE. The average between the two LHCb measurements is $\tau(\Lambda^0_b)^{\rm LHCb} = 1.468 \pm 0.009\, \mathrm{(stat.)} \pm 0.008\, \mathrm{(syst.)}$ ps.

\paragraph{$\Xi^-_b$, $\Xi^0_b$ and $\Omega^-_b$ lifetimes with $\mathcal{L}=3\ \mathrm{fb}^{-1}$}

The most abundantly produced $b$ baryon and therefore the most studied is the $\Lambda^0_b$. Less information are available for baryons that contain a strange quark and in particular only few and very weak theoretical predictions are available \cite{Lenz}. Thus, experimental results are useful to provide more stringent constraints on the models. The lifetimes of the $\Xi^-_b$, $\Xi^0_b$ and $\Omega^-_b$ baryons have been determined in a data sample corresponding to an integrated luminosity of 3 $\mathrm{fb}^{-1}$. The $\Xi^-_b$ and $\Omega^-_b$ baryons are reconstructed using the $\Xi^-_b \to J/\psi \ \Xi^-$ and $\Omega_b^- \to J/\psi \ \Omega^-$ decay modes. These lifetimes are determined using an absolute measurement with a technique similar to the one used for the determination of the $H_b \to J/\psi X$ lifetimes. However, due to the fact that the statistical uncertainty is greater in these two channels, the correction for the decay-time acceptance is extracted from simulated samples. The $\Xi^-_b$ and $\Omega^-_b$ baryon lifetimes are measured to be \mbox{$\tau(\Xi^-_b) = 1.55^{+0.10}_{-0.09}\, \mathrm{(stat.)} \pm 0.03\, \mathrm{(syst.)}$ ps} and $\tau(\Omega^0_b) = 1.54^{+0.26}_{-0.21}\, \mathrm{(stat.)} \pm 0.05\, \mathrm{(syst.)}$ ps \cite{baryon1}. A more recent measurement of the $\Xi_b^-$ lifetime using $\Xi_b^- \to \Xi_c^0 \pi^-$ decays, with $\Xi_c^0 \to p K^- K^- \pi^+$, leads to a value of \mbox{$\tau (\Xi^-_b) = 1.599 \pm 0.041\, \mathrm{(stat.)} \pm 0.018\, \mathrm{(syst.)} \pm 0.012\, \mathrm{(input)}$ ps} \cite{xibm_new}. 

The $\Xi^0_b$ baryon lifetime has been determined performing a relative \mbox{measurement \cite{baryon2}}. It is reconstructed using the $\Xi^0_b \to \Xi^+_c \pi^-$ decay mode and the lifetime is calculated relative to that of the $\Lambda^0_b$ baryon, reconstructed in the topologically similar \mbox{$\Lambda^0_b \to \Lambda^+_c \pi^-$} decay mode. From a fit to the background subtracted and efficiency corrected ratio of the yields as a function of the decay time, the ratio of the two lifetimes is determined to be $\tau(\Xi^0_b)/\tau(\Lambda^0_b) = 1.006 \pm 0.018\, \mathrm{(stat.)} \pm 0.010\, \mathrm{(syst.)}$. This shows that the $\Xi^0_b$ and the $\Lambda^0_b$ lifetimes are equal to within 2$\%$. Multiplying the lifetime ratio by $\tau(\Lambda^0_b)$ \cite{pdg}, the $\Xi^0_b$ baryon lifetime is determined to be $\tau(\Xi^0_b) = 1.477 \pm 0.026\, \mathrm{(stat.)} \pm 0.014\, \mathrm{(syst.)} \pm 0.013\, \mathrm{(input)}$ ps, where the last uncertainty is due to the precision of the $\Lambda^0_b$ lifetime. 

These are the most precise measurements to date and they lie in the range predicted by theoretical calculations.

\section{Conclusions}

\begin{figure}[htb]
\centering
\includegraphics[height=5cm]{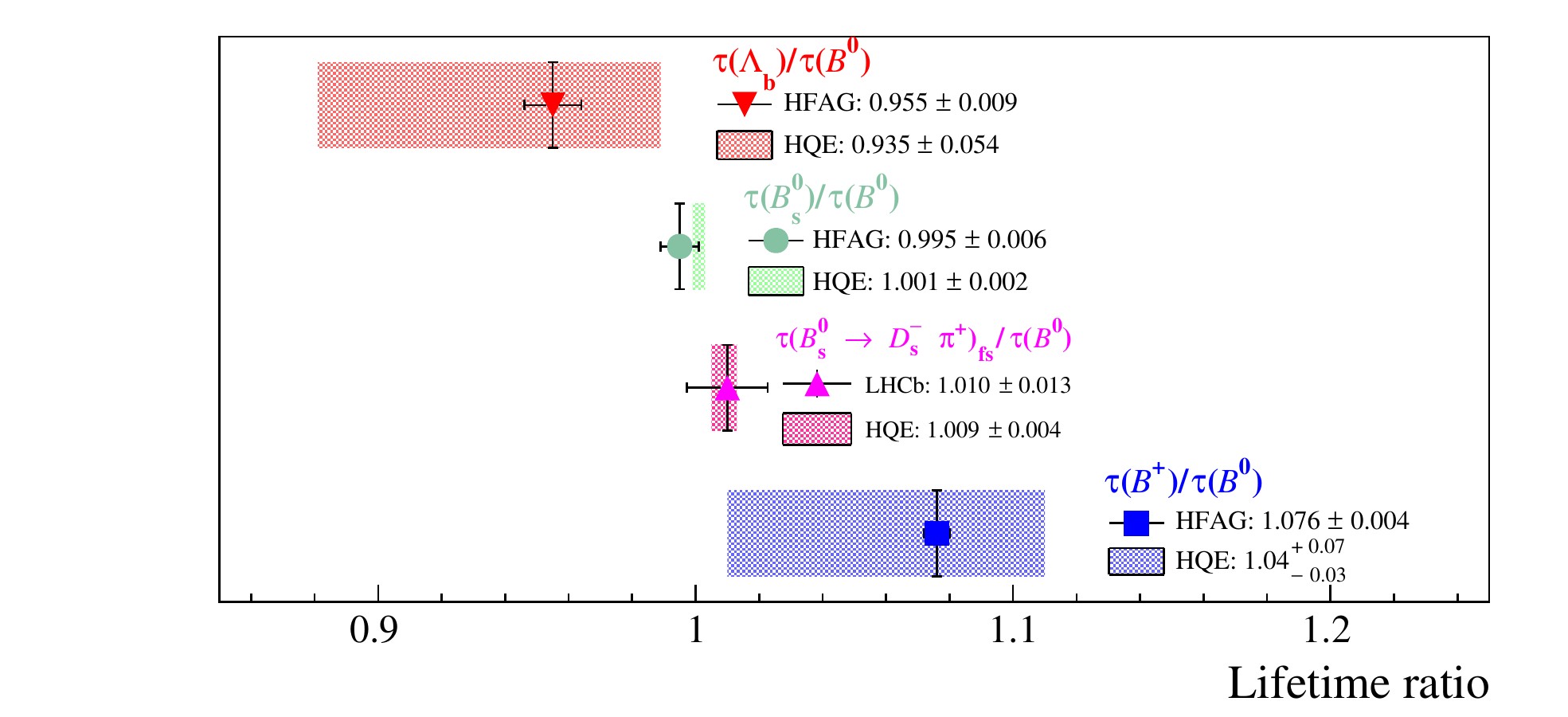}
\caption{Comparison of the LHCb lifetime ratios measurements reviewed in these proceedings with HQE predictions.}
\label{fig:comb}
\end{figure}

The LHCb lifetime measurements presented gives a consistent and quite complete picture of the different $b$-hadron lifetimes. 
All the lifetimes reported are the most precise measurements to date and are consistent with theoretical predictions~\cite{Lenz}, as it can be see in Figure~\ref{fig:comb}.
\begin{figure}[htb]
\centering
\includegraphics[height=5cm]{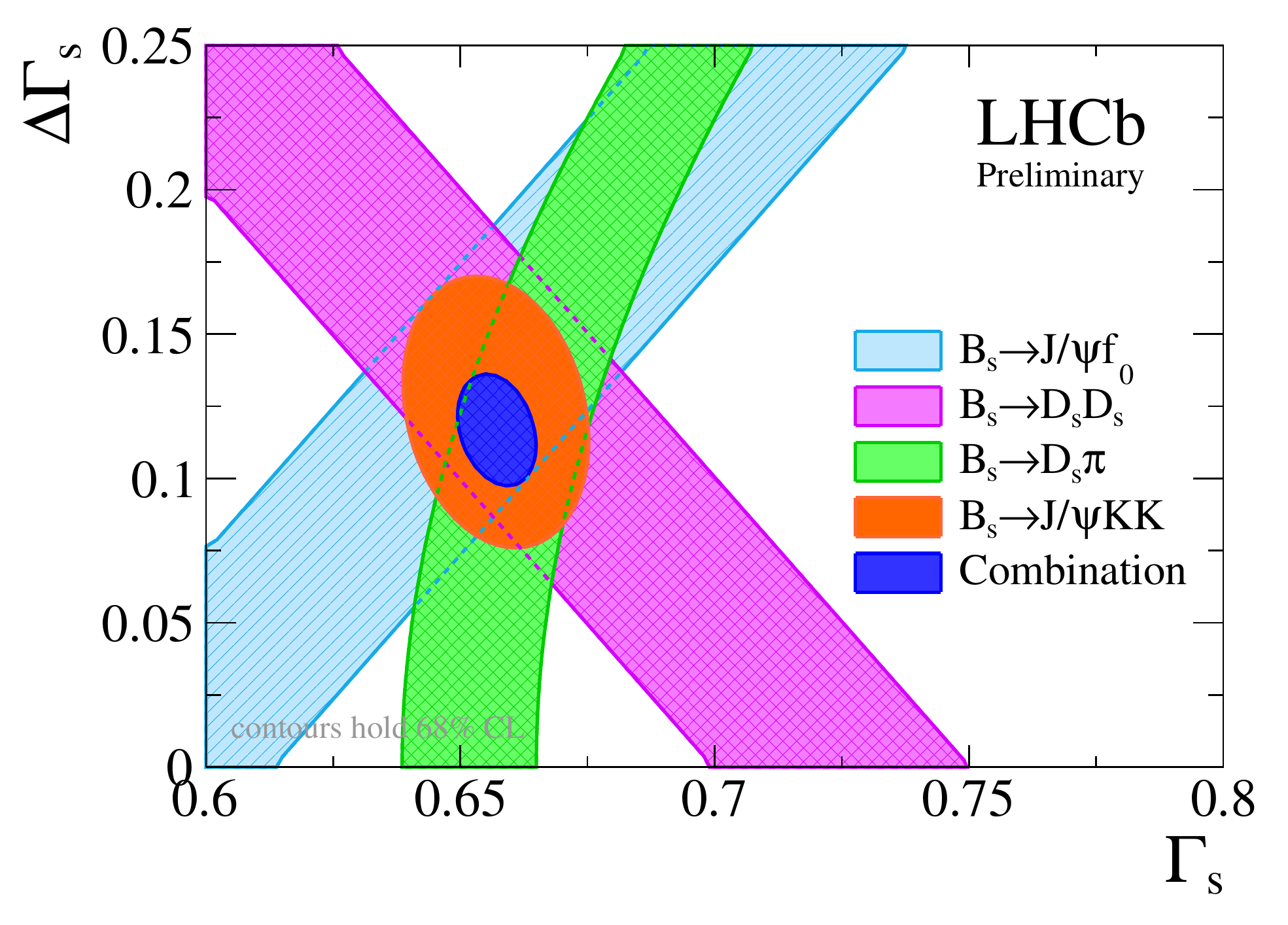}
\caption{Combination of LHCb $B^0_s$ effective lifetime results in ($\Gamma_s$, $\Delta \Gamma_s$) plane.}
\label{fig:comb_final}
\end{figure}

The $B^0_s$ effective lifetime measurements described in these proceedings have been included in the average shown in Figure~\ref{fig:comb_final}. It also includes the $B^0_s$ effective lifetime LHCb measurement determined in the final state $B^0_s \to J/\psi f_0$ \cite{lifetime1}. It shows the \mbox{$68\%$ CL} regions in the plane ($\Gamma_s$, $\Delta \Gamma_s$). The average taking all constraints into account is shown as the blue filled contour and it is in good agreement with the theory prediction $\Delta \Gamma_s = 0.087 \pm 0.021\, \mathrm{ps}^{-1}$~\cite{theory} (see also Ref. \cite{ut}), which assumes no new physics in $B^0_s$ mixing.

\Acknowledgements
The author acknowledges the support received from the ERC under FP7 and the support by
the International Max Planck Research School for Precision Tests of Fundamental Symmetries.

\end{document}